\documentclass[hidelinks,aps,pra,twocolumn,superscriptaddress,10pt]{revtex4-1} 

\usepackage{amsmath,amssymb}
\usepackage{mathrsfs}
\usepackage{epstopdf}
\usepackage{graphicx}
\usepackage{bm,color,bbm}
\usepackage{natbib}
\usepackage{hyperref}
\usepackage{caption}
\usepackage{subcaption}

\begin{document}


\title{Quantifying high-dimensional entanglement with Einstein-Podolsky-Rosen correlations}


\author{James Schneeloch}
\email{james.schneeloch@gmail.com}
\affiliation{Air Force Research Laboratory, Information Directorate, Rome, New York, 13441, USA}
\author{Gregory A. Howland}
\affiliation{Air Force Research Laboratory, Information Directorate, Rome, New York, 13441, USA}



\date{\today}

\begin{abstract}
Quantifying entanglement in a quantum system generally requires a complete quantum tomography followed by the NP-hard computation of an entanglement monotone--- requirements that rapidly become intractable at higher dimensions. Observing entanglement in large quantum systems has consequently been relegated to \emph{witnesses} that only verify its existence. In this article, we show that the violation of recent entropic witnesses of the Einstein-Podolsky-Rosen paradox also provides tight lower bounds to multiple entanglement measures, such as the \emph{entanglement of formation} and the \emph{distillable entanglement}, among others. Our approach only requires the measurement of correlations between two pairs of complementary observables---not a tomography---so it scales efficiently at high dimension. Despite this, our technique captures almost all the entanglement in common high-dimensional quantum systems, such as spatially or temporally entangled photons from parametric down-conversion.
\end{abstract}


\maketitle

Large-dimensional quantum systems shared by two distant parties are a key element of quantum protocols for secure communication and distributed or modular computing. Two individuals, Alice and Bob, might share remotely entangled qubit ensembles, or they may each possess a particle from a bipartite system entangled in high-dimensional degrees of freedom. Examples of the former include arrays of trapped ions \cite{Brown2016} or superconducting circuits \cite{Narla2016}, while latter examples include photon pairs entangled in their spatial or temporal variables \cite{Howell2004, Mair2001, Ali2007}. Because real-world systems are not ideal, it is necessary to characterize and quantify entanglement before it can be used as a quantum resource. 

Unfortunately, standard techniques for quantifying entanglement scale intractably at large dimensions, with optimum algorithms still being usable only for low-dimensional systems \cite{RyuEntCalc,AllendeEntCalc}. One must first perform a complete quantum tomography of the joint quantum system, followed by an NP-hard calculation of an entanglement monotone \cite{HuangNP}--- a function of a quantum state that decreases with entanglement-consuming operations, serving to measure entanglement. Two of the most popular entanglement monotones are the entanglement of formation $E_{F}$ and the distillable entanglement $E_{D}$. While the former indicates the number of two-qubit Bell states (known as \emph{ebits}) required on average to synthesize a copy of an entangled quantum state, the latter indicates the number of ebits that can be distilled on average out of a given quantum state. While both are entanglement monotones, it is straightforward to understand that $E_{F}\geq E_{D}$ \cite{christandl2004squashed}.

To mitigate the difficulty of calculating entanglement monotones, there are tight lower bounds to entanglement monotones using quantum entropy \cite{carlen2012bounds, AdabiTightEntUncRel2016, ChenEntBound}, but these bounds still require complete knowledge of the joint quantum state, which is still limited by the experimental difficulty of quantum state tomography. Even partial tomography of the joint density matrix allows one to certify entanglements of formation as high as 4.1 ebits in the temporal degree of freedom of photon pairs --- a world record set just last year \cite{MartinEntRecord}, though similar approaches have been proposed and tested elsewhere \cite{erker2015quantifying,bavaresco2017Quantifying}. Alternative approaches using optimizations of entanglement witness observables over their respective Hilbert spaces achieve even tighter lower bounds to entanglement monotones, but they become similarly difficult to apply to high-dimensional systems \cite{augusiak2009towards, guhne2009entanglement, brandao2005quantifying}.

Because of the infeasibility of high-dimensional tomography, and the difficulty of evaluating entanglement monotones, researchers interested in large-scale entangled systems have developed entanglement \emph{witnesses}. These only certify the presence of entanglement, but require dramatically fewer experimental and computational resources. Among these witnesses, there has been significant interest in \emph{entropic} entanglement witnesses of the Einstein-Podolsky-Rosen paradox \cite{EPR1935}, because they are directly applicable in quantum information protocols (where Shannon entropy characterizes both information and uncertainty), and because of their utility in one-sided device-independent secure communication. These witnesses, known as EPR-steering inequalities \cite{Wiseman2007, Cavalcanti2009}, are powerful because they make no assumptions about the state to be measured, whether pure or mixed, and whether Gaussian or wholly arbitrary; they only require the measurement of joint probability distributions of two pairs of complementary observables, and the calculation of conditional Shannon entropies from those distributions.

In this article, we show how the violation of these entropic EPR-steering inequalities \cite{Schneeloch2013} also provides large lower bounds to multiple entanglement monotones, directly connecting the extent of EPR-steering to the amount of entanglement. In particular, the negative quantum  conditional entropy, $-S(A|B)$ between two parties $A$ and $B$ forms a lower bound to the entanglement of formation $E_{F}$, and the violation of these EPR-steering inequalities allows us to  place lower bounds on the negative quantum conditional entropy, enabling us to readily quantify entanglement. For discrete observables, we show how Berta \emph{et~al}'s uncertainty principle in the presence of quantum memory \cite{Berta2010} can accomplish this. For continuous observables related by a Fourier transform (e.g., position $\hat{x}$ and momentum $\hat{k}=\hat{p}/\hbar$; energy $\omega=E/\hbar$ and time $t$, etc.) we develop the new relation:
\begin{equation}\label{OurResult}
h(x_{A}|x_{B})+h(k_{A}|k_{B})- \log(2\pi) \geq S(A|B)\geq -E_{F},
\end{equation} 
where $h(x_{A}|x_{B})$ is the continuous Shannon entropy of $x_{A}$ conditioned on $x_{B}$, determined by the joint probability density $\rho(x_{A},x_{B})$, and $S(A|B)$ is the conditional von Neumann (quantum) entropy of system $A$ conditioned on $B$, determined by joint density operator $\hat{\rho}_{AB}$. Here, and throughout this article, we measure entropy in bits, so that all logarithms are base 2. While demonstrating the EPR paradox has fundamental applications in one-sided device-independent quantum secure communication, these same correlations will now also be of great utility across all disciplines of quantum physics, wherever entanglement is utilized. Moreover, most sources of entangled pairs of particles have correlations well-captured by this variety of quantitative witness due to a common place of origin producing position (alt. time) correlations and conservation laws producing momentum (alt. energy) anti-correletions.

Our entanglement quantification scheme relies on two elements. First, the uncertainty principle in the presence of quantum memory \cite{Berta2010} allows us to relate the strength of correlations of complementary observables to the quantum conditional entropy. Given two $N$-dimensional observables $\hat{Q}$ and $\hat{R}$, and two parties $A$ and $B$, the uncertainty principle in the presence of quantum memory, using classical measurements, is given by:
\begin{equation}\label{Berta}
H(\hat{Q}_{A}|\hat{Q}_{B}) + H(\hat{R}_{A}|\hat{R}_{B}) \geq \log(\Omega_{QR}) + S(A|B).
\end{equation}
Here, $H(\hat{Q}_{A}|\hat{Q}_{B})$ is the conditional Shannon entropy of the discrete probability distribution $P(Q_{Ai},Q_{Bj})=\text{Tr}\big[\hat{\rho}_{AB}\;|q_{Ai}\rangle\langle q_{Ai}|\otimes | q_{Bj}\rangle\langle q_{Bj}|\big]$, and $|q_{Ai}\rangle$ is the $i^{\text{th}}$ eigenstate of $\hat{Q}_{A}$. The bound $\Omega_{QR}$ approaches the dimensionality $N$ of system $A$ (or $B$) when observables $\hat{Q}$ and $\hat{R}$ are mutually unbiased \cite{ColesRMP}.  When there are large correlations between $\hat{Q}_{A}$ and $\hat{Q}_{B}$ and between $\hat{R}_{A}$ and $\hat{R}_{B}$, the quantum conditional entropy $S(A|B)$ may be less than zero, witnessing entanglement between $A$ and $B$. The inequality \eqref{Berta} without the state-dependent term $S(A|B)$ is an entropic EPR-steering inequality whose violation both witnesses entanglement and demonstrates the EPR paradox \cite{Schneeloch2013}.

The second element of our entanglement quantification scheme uses recent relations between entanglement monotones and the negative quantum conditional entropy. In \cite{carlen2012bounds} (and also discussed in \cite{AdabiTightEntUncRel2016}), it is shown that the entanglement of formation $E_{F}$, the relative entropy of entanglement $E_{RE}$, and the squashed entanglement $E_{SQ}$ are all bounded below by the largest negative quantum conditional entropy:
\begin{equation}\label{rel1}
\{E_{F},E_{RE},E_{SQ}\}\geq\max\{0,-S(A|B),-S(B|A)\}
\end{equation}
Furthermore, in \cite{christandl2004squashed}, the distillable entanglement $E_{D}$ is bounded from below by the mean of both negative quantum conditional entropies:
\begin{equation}\label{rel2}
E_{D}\geq\frac{1}{2}(-S(A|B) -S(B|A)).
\end{equation}
Together, these two elements allow us to place large lower bounds on entanglement monotones using strong correlations across complementary observables. For entropic EPR-steering inequalities \cite{Schneeloch2013} for pairs of $N$-dimensional systems, the amount by which you violate the inequality gives a minimum value for the entanglement monotones $\{E_{F},E_{RE},E_{SQ}\}$, while the mean of the violations for EPR-steering in each direction gives a minimum value for $\{E_{D}\}$. Among other applications, this technique will be of great use in quantifying the entanglement between two large groups of qubits. In these situations, even state tomography becomes prohibitively difficult because the number of measurements needed for state tomography grows with the fourth power of each party's dimension.

To adapt \eqref{Berta} into an inequality for continuous observables, we define a pair of discrete observables related by a quantum Fourier transform that converge to a pair of continuous observables related by a continuous Fourier transform in the appropriate limit. Similar approaches introducing continuous position and momentum as a limit of discrete observables may be found in \cite{Shankar,CresserNotes}, while a similar approach deriving a continuous variable (qualitative) entanglement witness as a limit of discrete uncertainty relations can be found in \cite{tasca2013reliable}.

Let us define two $N$-dimensional observables $\hat{X}$ and $\hat{K}$ such that:
\begin{equation}
\hat{X}\equiv\sum_{\ell=-N/2}^{N/2-1} X_{\ell} |X_{\ell}\rangle\langle X_{\ell}|,
\end{equation}
where,
\begin{equation}
X_{\ell}\equiv \ell\cdot\Delta x,
\end{equation}
and such that
\begin{equation}
\hat{K}\equiv\sum_{m=-N/2}^{N/2-1} K_{m} |K_{m}\rangle\langle K_{m}|,
\end{equation}
where,
\begin{equation}
K_{m}\equiv m\cdot\frac{2\pi}{N\Delta x},
\end{equation}
and the relation between $\hat{X}$ and $\hat{K}$ is that of a quantum Fourier transform:
\begin{equation}
\langle X_{\ell}|K_{m}\rangle\equiv\frac{1}{\sqrt{N}}e^{i\frac{2\pi}{N}\ell m}.
\end{equation}

The uncertainty principle in the presence of quantum memory for measurements of $\hat{X}$ and $\hat{K}$ takes the form:
\begin{equation}\label{xpdisc}
H(\hat{X}_{A}|\hat{X}_{B}) + H(\hat{K}_{A}|\hat{K}_{B})\geq\log\Big(\frac{2\pi}{\Delta x_{A}\Delta k_{A}}\Big)+S(A|B),
\end{equation}
where we express $N$ in terms of our defined $\Delta x$ and $\Delta k$:
\begin{equation}
\Delta x\Delta k=\frac{2\pi}{N}.
\end{equation}

\begin{figure}[t]
\centerline{\includegraphics[width=\columnwidth]{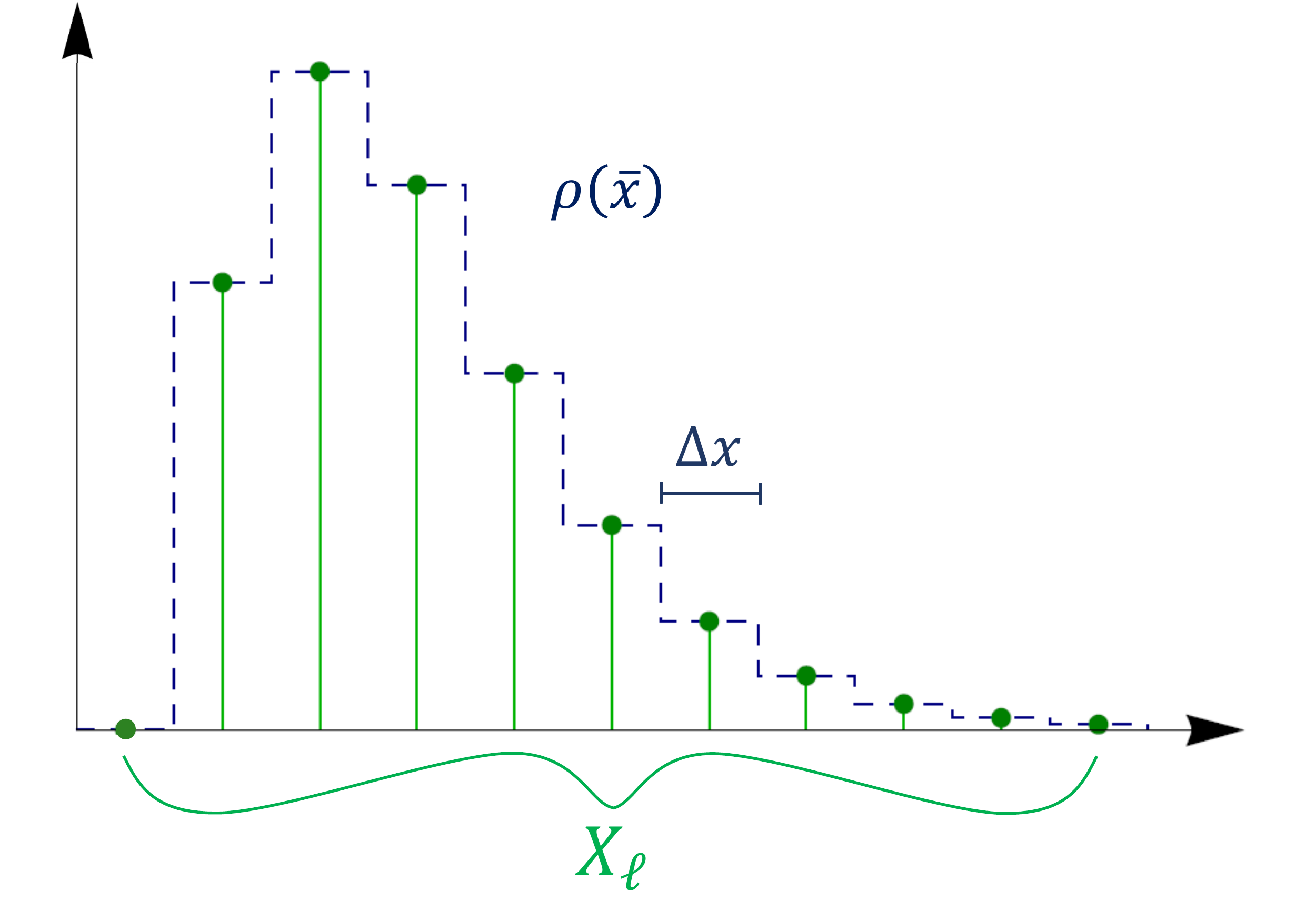}}
\caption{Diagram of continuous-variable analogue probability density $\rho(\bar{x})$ (dotted line) to discrete probability distribution $P(X_{\ell})$. The green vertical lines indicate the values of $X_{\ell}$ in the discrete probability distribution $P(X_{\ell})$.}
\end{figure}

Next, we define continuous - variable analogues $(\bar{x}_{A},\bar{x}_{B},\bar{k}_{A},\bar{k}_{B})$ to the discrete variables $(\hat{X}_{A},\hat{X}_{B},\hat{K}_{A},\hat{K}_{B})$. The probability densities for $(\bar{x}_{A},\bar{x}_{B},\bar{k}_{A},\bar{k}_{B})$ are quantized into bins of size $(\Delta x_{A},\Delta x_{B}, \Delta k_{A}, \Delta k_{B})$ such that integrating over these bins gives the corresponding probabilities of the discrete variables $(\hat{X}_{A},\hat{X}_{B},\hat{K}_{A},\hat{K}_{B})$. See Fig.~1 for diagram. 

The continuous Shannon entropy is defined as a limit of the discrete Shannon entropy plus a distribution-independent offset \cite{Cover2006}. Because the discrete variables $(\hat{X}_{A},\hat{X}_{B},\hat{K}_{A},\hat{K}_{B})$ obey the inequality \eqref{xpdisc}, and the continuous analogue variables $(\bar{x}_{A},\bar{x}_{B},\bar{k}_{A},\bar{k}_{B})$ obey the relation:
\begin{equation}
H(\hat{X}_{A}|\hat{X}_{B}) + \log(\Delta x_{A}) = h(\bar{x}_{A}|\bar{x}_{B}),
\end{equation}
they must also obey the inequality:
\begin{equation}\label{approx}
h(\bar{x}_{A}|\bar{x}_{B}) + h(\bar{k}_{A}|\bar{k}_{B})\geq \log(2\pi) +S(A|B).
\end{equation}
 
This inequality \eqref{approx} is valid no matter the values of $\Delta x_{A}$, $\Delta x_{B}$, $N_{A}$ or $N_{B}$. In the limit as $N_{A}$ and $N_{B}$ approach $\infty$ followed by the limit as $\Delta x_{A}$ and $\Delta x_{B}$ approach zero, the discrete observables $(\hat{X}_{A},\hat{X}_{B},\hat{K}_{A},\hat{K}_{B})$ approach observables with unbounded continuous spectra related by continuous Fourier transforms. These limiting observables are identical to continuous position $\hat{x}$ and momentum $\hat{k}$. See Supplemental material for details. Consequently, the probability densities for the resulting continuous observables $(\hat{x}_{A},\hat{x}_{B},\hat{k}_{A},\hat{k}_{B})$ must obey the same inequality. This proves our first relation. For particles $A$ and $B$, the probability densities for continuous position observables $(\hat{x}_{A},\hat{x}_{B})$, and momentum observables $(\hat{k}_{A},\hat{k}_{B})$ are constrained by the inequality:
\begin{equation}\label{result}
h(x_{A}|x_{B}) + h(k_{A}|k_{B})\geq\log(2\pi)+S(A|B)
\end{equation}
This inequality solves one aspect of the important open problem posed by \cite{furrer2014position} --- how to extend the uncertainty principle in the presence of quantum memory to continuous observables, at least for classical measurements. For newly defined varieties of quantum continuous conditional entropy, this has been accomplished \cite{Frank2012Entropy}, but fully quantum relations require full quantum tomography to be applied. Our result \eqref{result} will have far-reaching implications in experimental investigations of quantum entanglement, as we explore further in this article.

The techniques we used to develop our continuous-variable entropic uncertainty relation \eqref{result} apply to any pair of Fourier-conjugate observables. While we derived our relation for continuous position and momentum, they apply equally well for conjugate field quadratures $\hat{u}=\frac{1}{\sqrt{2}}(\hat{a}^{\dagger} + \hat{a})$ and $\hat{v}=\frac{i}{\sqrt{2}}(\hat{a}^{\dagger}-\hat{a})$, with commutator $[\hat{u},\hat{v}]=i$:
\begin{equation}\label{squeezing}
h(u_{A}|u_{B}) + h(v_{A}|v_{B})\geq \log(2\pi) + S(A|B)
\end{equation}
Using the more common convention for field quadratures where $\hat{u}=\frac{1}{2}(\hat{a}^{\dagger} + \hat{a})$ and $\hat{v}=\frac{i}{2}(\hat{a}^{\dagger}-\hat{a})$, with commutator $[\hat{u},\hat{v}]=i/2$ only changes the relation by adding 1 bit to the bound due to the scaling law of continuous entropies.

Just as with continuous position and momentum, it is also straightforward to show how our technique works for (bounded continuous) angular position $\theta$ and (unbounded discrete) angular momentum $\ell_{z}=L_{z}/\hbar$:
\begin{equation}
H(\ell_{zA}|\ell_{zB}) + h(\theta_{A}|\theta_{B})\geq \log(2\pi) + S(A|B),
\end{equation}
One needs only to take the limits $\Delta x\rightarrow 0$ and $N\rightarrow\infty$ while holding the product $N\Delta x$ contant, to hold $\Delta k$ constant (or equivalently $\Delta\ell_{z}$). Consequently for photon number $N$ and phase $\phi$, which are also Fourier-conjugate observables of a single mode of the quantized electromagnetic field \cite{ColesRMP}:
\begin{equation}
H(N_{A}|N_{B}) + h(\phi_{A}|\phi_{B})\geq \log(2\pi) + S(A|B).
\end{equation}

Moreover, since angular frequency $\omega$ and (relative) time $t$ are also Fourier-conjugates, we also obtain the analogous inequality relating frequency and (arrival) time correlations:
\begin{equation}
h(t_{A}|t_{B}) + h(\omega_{A}|\omega_{B})\geq \log(2\pi) + S(A|B).
\end{equation}
These new relations allow straightforward quantification of high-dimensional entanglement using orbital angular momentum (OAM) measurements, quadrature measurements, frequency and time measurements in addition to transverse spatial measurements. These degrees of freedom are all promising candidates for high-density quantum communication.

Furthermore, while the original discrete observables $\hat{X}_{A}$ and $\hat{K}_{A}$ in \eqref{xpdisc} are related by a quantum Fourier transform, the variables $\hat{X}_{B}$ and $\hat{K}_{B}$ are wholly arbitrary and may correspond to observables in disparate degrees of freedom (e.g., polarization). This affords us the capability of quantifying hybrid entanglement as well. For polarization observables $\hat{\sigma}_{XB}$ and $\hat{\sigma}_{ZB}$, the following inequality:
\begin{equation}
H(\hat{X}_{A}|\hat{\sigma}_{XB}) + H(\hat{K}_{A}|\hat{\sigma}_{ZB})\geq \log\Big(\frac{2\pi}{\Delta x_{A}\Delta k_{A}}\Big) + S(A_{x}|B_{\sigma})
\end{equation}
will witness entanglement between a spatial degree of freedom of particle $A$ (represented by $A_{x}$), and the spin degree of freedom for particle $B$ (represented by $B_{\sigma}$).

Additionally, in the limit that particles $A$ and $B$ are completely independent of one another, our relation \eqref{result} reduces to the position-momentum uncertainty relation for a single particle including quantum entropy discussed in \cite{Frank2012Entropy}, i.e.,
\begin{equation}
h(x_{A}) + h(k_{A}) \geq \log(2\pi) + S(A).
\end{equation}

To adapt our continuous-variable uncertainty relation \eqref{result} to experimental measurements carried out at finite resolution, we follow the same procedure used in \cite{Schneeloch2012,Schneeloch2013relationship} to adapt Walborn \emph{et al}'s continuous-variable EPR-steering inequality to discrete measurements. Simply put, coarse graining of any kind is a non-decreasing operation on continuous entropy. Consequently, the inequality:
\begin{equation}
H(X_{A}|X_{B}) + H(K_{A}|K_{B})\geq\log\Big(\frac{2\pi}{\Delta x_{A}\Delta k_{A}}\Big)+S(A|B).
\end{equation}
is valid for position and momentum measurements carried up to (now arbitrary and independent) resolutions $\Delta x$ and $\Delta k$, respectively.

As one additional improvement, we may add together bounds for the entanglement between respective degrees of freedom to obtain a bound for the total entanglement. This comes from the sub-additivity of the quantum conditional entropy, e.g.,:
\begin{equation}
S(A_{x},A_{\sigma}|B_{x},B_{\sigma})\leq S(A_{x}|B_{x}) + S(A_{\sigma}|B_{\sigma}).
\end{equation}
With this, we need only add the lower bounds in each degree of freedom to obtain the total lower bound for the entanglement of formation (among other measures) of the complete joint state.

\begin{figure}[t]
\centerline{\includegraphics[width=\columnwidth]{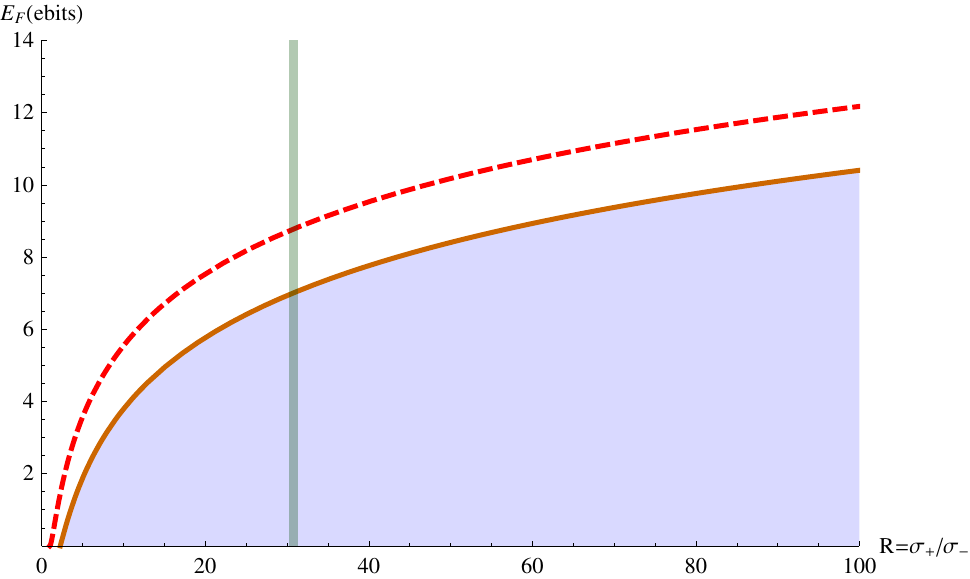}}
\caption{Plot comparing the entanglement of formation of a double-Gaussian state in both transverse dimensions (dotted curve) to the amount of entanglement our inequality witnesses as a function of the correlation ratio $R$ (solid curve). The difference between the two rapidly approaches a constant value of 1.771 ebits. The vertical bar indicates the correlation ratio $R\approx 30.8$ predicted for Type-I SPDC using a 390 nm Gaussian pump beam with a 0.68 mm beam diameter incident on a 2.0 mm BiBO nonlinear crystal.}
\end{figure}

With our inequality \eqref{result}, we can efficiently establish the presence of a large amount of entanglement. To give an idea of how much entanglement we can witness, we can obtain a looser, variance-based relation in the same fashion as Heisenberg's relation is a limit of Bialynicki-Birula and Mycielski's entropic relation. In particular, it is straightforward to show that:
\begin{equation}
h(x_{A}|x_{B})\leq h(x_{A}\pm x_{B}) \leq \frac{1}{2}\log\Big(2\pi e \sigma^{2}_{(x_{A}\pm x_{B})}\Big),
\end{equation}
and from this, to get the quantitative entanglement witness for position and momentum variances:
\begin{equation}
E_{F}\geq -\log\Big(e\; \sigma_{(x_{A}\pm x_{B})}\sigma_{(k_{A}\mp k_{B})}\Big)
\end{equation}
Similar relations can be used to witness entanglement using variances in field quadratures, frequencies, and other continuous-variable degrees of freedom.

Using this variance-based relation, we discover that the spatial correlations measured in one of the seminal papers on spatial entanglement \cite{Howell2004} thirteen years ago, actually indicate at least 1.88 ebits of spatial entanglement (their value of $\sigma_{(x_{A} - x_{B})}\sigma_{(k_{A}+ k_{B})}$ is about $0.1$), though the witnessed entanglement would likely be doubled if they measured in both transverse dimensions, and increased further with higher experimental resolution. Indeed, later experiments measuring these variances in similar systems quantify as much as $3.88$ ebits of spatial entanglement \cite{Edgar2012}. While this alone doesn't break the record set by Martin et al last year, it does break the same record they do, showing that our relations witness comparable amounts of entanglement with much simpler measurement schemes.

To examine the efficiency of our quantitative entanglement witness (i.e., how much of the total entanglement can be verified), we consider one of the only entangled wavefunctions whose entanglement of formation can be explicitly evaluated to have a basis of comparison. The double-Gaussian state approximates the transverse spatial correlations of photon pairs generated by Spontaneous Parametric Down-Conversion (SPDC) (See Fig.~2) \cite{LawEberly2004, Schneeloch_SPDC_Reference_2016}. For a pair of particles $A$ and $B$, in a double-Gaussian state, their joint position wavefunction $\psi(x_{A},x_{B})$ is given by:
\begin{equation}
\psi(x_{A},x_{B})=\frac{1}{\sqrt{2\pi\sigma_{+}\sigma_{-}}}e^{-\frac{(x_{A}+x_{B})^{2}}{8\sigma_{+}^{2}}}e^{-\frac{(x_{A}-x_{B})^{2}}{8\sigma_{-}^{2}}}.
\end{equation}
This wavefunction closely approximates the state originally considered in the EPR paradox.

For the Double-Gaussian state (in one transverse dimension), the sum of position and momentum conditional entropies is given by:
\begin{equation}
h(x_{A}|x_{B})+h(k_{A}|k_{B})=\log\Big(\frac{2\pi e}{R + 1/R}\Big)
\end{equation}
where $R$ is the (correlation) ratio of the standard deviations $\sigma_{(x_{A}+x_{B})}$ over $\sigma_{(x_{A}-x_{B})}$. For the experimental system in \cite{Howell2004}, which also uses SPDC, with a 390~nm pump beam with a $1/e^{2}$ beam diameter of 0.68~mm, using a 2.0~mm long nonlinear crystal, this ratio $R\approx 30.8$, indicating an entanglement of formation (using both transverse dimensions) greater than or equal to $7.00$ ebits. While the Double-Gaussian state  only approximates the transverse spatial wavefunction of entangled photon pairs, it is the state with maximum entanglement for a given correlation ratio $R$ \cite{Schneeloch_SPDC_Reference_2016}, so that the actual entanglement of formation of a state with these correlations is at most $8.77$ ebits (See Fig. 2 and supplementary material for details). In order to simply witness entanglement, correlation ratios as small as 2.28 are sufficient for our quantitative witness. These bounds indicate our techniques can witness within two ebits of the total entanglement present in the system in two dimensions, or within one ebit in one dimension. This may yet be further improved to quantify all the spatial entanglement, as the discrepancy between the witnessable entanglement of the Double-Gaussian state, and its actual entanglement of formation is directly related to the difference between our inequality's \eqref{result} fixed bound of $\log(2\pi)$ and the bound of $\log(e\pi)$ for the corresponding position-momentum EPR-steering inequality \cite{Walborn2011}.

Here we have developed a new technique to verify large amounts of entanglement using only the correlations needed to demonstrate the EPR paradox. In particular, the amount by which entropic EPR-steering inequalities \cite{Schneeloch2013} can be violated is directly related to the amount of entanglement that can be verified. With standard experimental equipment, one may witness over 10 ebits of spatial entanglement in SPDC photon pairs. As a direction for future investigation, we point out that the frequency/time correlations between photon pairs in SPDC are theoretically strong enough to witness in excess of 30 ebits of entanglement (e.g., by pumping a 0.5mm BiBO crystal at 775nm to produce spectrally degenerate SPDC with a narrow linewidth laser) \cite{Schneeloch_SPDC_Reference_2016}). While 30 ebits represents more entanglement than what a quintillion-dimensional (or 60-qubit) joint state can support, techniques to fully resolve these correlations remain to be developed.

\begin{acknowledgments}
We gratefully acknowledge support from the National Research Council Research Associate Programs, and funding from the OSD ARAP QSEP program, as well as insightful discussions with Dr. Paul M. Alsing, Mr. Michael L. Fanto, and Dr. Christopher C. Tison. Any opinions, findings and conclusions or recommendations expressed in this material are those of the author(s) and do not necessarily reflect the views of AFRL.
\end{acknowledgments}

\bibliography{EPRbib16}

\newpage

\appendix

\section{Supplementary Material}
\subsection{The Continuum Limit}
To facilitate the continuum limit, we define the non-normalized eigenkets of $\hat{X}$ and $\hat{K}$:
\begin{equation}
|\tilde{X}_{\ell}\rangle\equiv\frac{|X_{\ell}\rangle}{\sqrt{\Delta x}}\qquad;\qquad|\tilde{K}_{m}\rangle\equiv\frac{|K_{m}\rangle}{\sqrt{\Delta k}}
\end{equation}
Expressing $\hat{X}$ and $\hat{K}$ in terms of these non-normalized eigenkets gives us, e.g.,
\begin{equation}
\hat{X}=\sum_{\ell=-N/2}^{N/2-1}\Delta x \;\;X_{\ell} |\tilde{X}_{\ell}\rangle\langle \tilde{X}_{\ell}|
\end{equation}
and a similar expression for $\hat{K}$.

The continuum limit is taken where first, $N$ approaches infinity, and then $\Delta x$ approaches zero. In this sequence of limits, the sums become integrals, where:
\begin{equation}
\hat{X}\rightarrow\int_{-\infty}^{\infty}dx_{\ell} \;\;x_{\ell} |\tilde{x}_{\ell}\rangle\langle \tilde{x}_{\ell}|=\hat{x}
\end{equation}
and $\hat{K}$ approaches its corresponding integral. Taking the first limit (i.e., $N\rightarrow\infty$), $\hat{X}$ becomes an observable with a discrete, but unbounded spectrum, with eigenvalues evenly spaced apart by amount $\Delta x$, while $\hat{K}$ approaches an observable with a bounded but continuous spectrum, since in this limit $\Delta k\rightarrow 0$, but $N\Delta k$ remains constant. Then,  taking the second limit $\Delta x\rightarrow 0$, the spectrum of $\hat{X}$ becomes both continuous and unbounded, while for $\hat{K}$ the bounds to the spectrum approach infinity (making the ``momentum'' spectrum unbounded as well. Thus, in this limit, $\hat{X}$ and $\hat{K}$ approach observables with unbounded continuous spectra still related by a quantum Fourier transform. Because of our choice of definition for the eigenvalues $K_{m}$, the relation between $\hat{X}$ and $\hat{K}$ becomes that for continuous position and momentum:
\begin{equation}
\langle\tilde{X}_{\ell}|\tilde{K}_{m}\rangle=\frac{1}{\sqrt{2\pi}}e^{i X_{\ell}K_{m}}\rightarrow\langle x|k\rangle.
\end{equation}
Since our defined observables approach continuous observables related by a continuous Fourier Transform as with continuous position and momentum, we can apply these limits to \eqref{Berta} to obtain our relation.

\subsection{Relation for Quantum Conditional Entropy}
The relation:
\begin{equation}
S(A_{x},A_{\sigma}|B_{x},B_{\sigma})\leq S(A_{x}|B_{x}) + S(A_{\sigma}|B_{\sigma}).
\end{equation}
can be re-expressed in terms of quantum mutual information:
\begin{equation}
I(B_{x}:B_{\sigma})\leq I(A_{x},B_{x}:A_{\sigma},B_{\sigma}).
\end{equation}
The relation is valid because adding new systems cannot decrease the quantum mutual information.
\newline

\subsection{Maximum Entanglement of Formation of a Biphoton Wavefunction for a given Correlation Ratio}
As mentioned previously, the Double-Gaussian state is the maximally entangled state for a given correlation ratio $R$. To show this, we point out that the Schmidt eigenvalues of the Double-Gaussian state form a maximum entropy probability distribution for a given mean, (i.e., a geometric distribution), and this entropy is equal to the entanglement of formation. Second, the mean value of this probability distribution is a one-to-one function of the correlation ratio $R$ for all values of $R\geq 1$. With this, the Schmidt eigenvalues of the double-Gaussian state are also a maximum entropy probability distribution for a given value of $R$, so long as $R\geq 1$. Therefore, a given biphoton wavefunction, with a given correlation ratio $R$ will have a maximum entanglement of formation of what a double-Gaussian state would have for the same value of $R$:
\begin{equation}
E_{f}(\hat{\rho}_{AB})\leq\frac{h_{2}(\lambda)}{\lambda}\qquad:\qquad\lambda\equiv\frac{4R}{(R+1)^{2}}.
\end{equation}
Here, $h_{2}(\lambda)$ is the binary entropy function:
\begin{equation}
h_{2}(\lambda)=-\lambda \log_{2}(\lambda) - (1-\lambda)\log_{2}(1-\lambda).
\end{equation}

\end{document}